\documentclass[12pt,a4paper]{article}
\usepackage{amsfonts,amssymb,amsmath,amsthm,cite}
\usepackage{graphicx}
\usepackage{slashed}

\newcommand{\beq}{\begin{eqnarray*}}
\newcommand{\eeq}{\end{eqnarray*}}

\newcommand{\ba}{\begin{array}}
\newcommand{\ea}{\end{array}}

\newcommand{\Tr}{\mathrm{Tr}\, }
\newcommand{\ii}{\mathrm{i}}
\newcommand{\bbeta}{\tau}

\begin{document}
\title{\bf Soliton Fermionic number \\ from \\ the heat kernel expansion
 }
\author{A. Alonso-Izquierdo$^{1}$\footnote{alonsoiz@usal.es}, Rodrigo Fresneda$^{2}$\footnote{rodrigo.fresneda@ufabc.edu.br},\\ J. Mateos Guilarte$^{3}$\footnote{guilarte@usal.es},, D. Vassilevich$^{2,4}$\footnote{dvassil@gmail.com}
\\
\small $^{1}$ Departamento de Matematica Aplicada, University of Salamanca, Spain\\
\small $^{2}$ CMCC, Universidade Federal do ABC, Santo Andre, Sao Paulo, Brazil\\
\small $^{3}$  Departamento de Fisica Fundamental, University of Salamanca, Spain \\
\small $^{4}$ Physics Department, Tomsk State University, Tomsk, Russia
}

\date{\today}

\maketitle
\begin{abstract}
We consider different methods of calculating the (fractional) fermion number of solitons based on the heat kernel expansion. We derive a formula for the localized $\eta$ function that provides a more systematic version of the derivative expansion for spectral asymmetry and compute the fermion number in a multiflavor extension of the Goldstone-Wilczek model. We also propose an improved expansion of the heat kernel that allows the tackling of the convergence issues and permits an automated computation of the coefficients.
\end{abstract}

\section{Introduction}
More than 40 years ago Jackiw and Rebbi \cite{Jackiw:1975fn} discovered that the Fermi number of solitonic ground states may be half-integer. Later, it was demonstrated that fractional and even  irrational values of the fermion number are also allowed, see \cite{Goldstone:1981kk} and the review paper \cite{Niemi:1984vz}, where the whole structure of fermion fractionization by solitons was exhibited
for several systems with solitons of different dimensions and properties, namely, 1D kinks, planar vortices, and 3D magnetic monopoles. A crucial step in understanding the fermion fractionization in solitonic backgrounds was given by Niemi and Semenoff in Reference \cite{Niemi:1983rq} where it was reinterpreted in the context of QFT anomalies.

On the theoretical side this unexpected phenomenon showed that very subtle mathematical concepts arising in topological index theory played an important r$\hat{\rm o}$le in its understanding. Together with other phenomena, anomalies, instantons, etcetera, fermionic fractionization triggered by solitons laid the foundations for the extraordinary cross fertilization between
Mathematics and Quantum Physics within the QFT framework successfully developed during the last forty years.

On the experimental side Fermion fractionization meant electric charge fractionization violating the Millikan principle observed in Nature, regarding the indivisibility of the electron charge. This possibility is susceptible of being experimentally tested and a positive result on
 the existence of electric charges not being an integer multiple of the electron charge was found in
exotic macromolecules, e.g. polyacetylene, see \cite{Heeger:1979prl,Su:1984}. The link between the existence of the fermion fractionization phenomenon
in condensed matter physics and relativistic field theory was established in Reference \cite{Schrieffer:1981} giving rise to one more connection between these two
apparently distant areas of Physics. Also, much of the renewed interest in this problem was caused by various applications to Condensed Matter Physics, see e.g. the papers \cite{Jackiw:2007rr,Chamon:2007hx} that treat the charge fractionization in graphene-related models. Various condensed matter applications bring up more and more complex systems \cite{Wang:2017xhg,Lepori:2018vwg,Martin-Ruiz:2018tdv} that require more and more sophisticated methods of analysis.

Most reliable methods use anomalies to compute the charge fractionization. Such methods however cannot be applied to all models. One thus needs certain kinds of the derivative expansion \cite{Goldstone:1981kk,Chamon:2007hx}. This method is however not free of problems. On solitonic backgrounds the fields are usually of the same order as their derivatives. Thus application of the derivative expansion to Feynman diagrams is tricky, besides the combinatorial complexity if matrix-valued fields are involved. On top of this, convergence of the expansion is far from being obvious. {\footnote{We also like to mention at this point that the related trace identities method was successfully applied in \cite{Paranjape} to the computation of fermi fractionization for magnetic monopoles.}}

The purpose of this work is to reconsider and improve the calculation methods of the fractional fermion number of solitons by systematically using heat kernel techniques (see, e.g., \cite{Vassilevich:2003xt} for a review). We shall concentrate on combinatorial aspect, derive a new expression for the fractional fermion number in a model with matrix-valued fields, and analyze the convergence issues in an improved version of the heat kernel expansion.

Note, that for computations of the parity anomaly \cite{Niemi:1983rq,Redlich:1983kn}, which is a close cousin of the problem considered here, the heat kernel methods proved to be very efficient \cite{AlvarezGaume:1984nf,Deser:1997nv,Deser:1997gp,Kurkov:2017cdz,Kurkov:2018pjw}. The same can be said about the mass shift of solitons due to quantum corrections. The usual heat kernel expansion alone delivers the answer in the supersymmetric case \cite{Bordag:2002dg,Vassilevich:2003xk}. In non-supersymmetric case, however, certain modifications of the expansion are needed \cite{AlonsoIzquierdo:2002eb}.

Our strategy is as follows. In the next section we derive localized expressions for the $\eta$ function and for the spectral asymmetry. This allows us to use the standard heat kernel expansion and derive an expression for the fermion number in a non-abelian (multiflavor) generalization of the Goldstone-Wilczek model. Then to address certain drawbacks of the standard heat kernel expansion we develop an improved version of the heat kernel. To set up the stage, we re-derive the Niemi-Semenoff formula for fermion number. In Section \ref{sec:eta} we analyze convergence of the expansions.

\section{Fermion number through the heat kernel expansion}\label{sec:hk}
Let us consider a theory of Dirac fermions in $d$ spatial dimensions described by a Hamiltonian $H$.
The fermion number $N$ is given by the formula \cite{Niemi:1984vz}
\begin{equation}
N= -\tfrac 12 \eta(0,H), \label{eq:N}
\end{equation}
where $\eta (0,H)$ is the value at $0$ of the spectral $\eta$ function that measures the spectral asymmetry of the Dirac Hamiltonian $H$,
\begin{equation}
\eta (s,H)=\sum_{\lambda >0} \lambda^{-s} - \sum_{\lambda <0} (-\lambda)^{-s}. \label{zeta0H}
\end{equation}
Here $\lambda$'s are the eigenvalues of $H$. Through a Mellin transform, this formula can be converted to
\begin{equation}
\eta (s,H)= \Tr \left(  (H^2)^{-s/2} H/|H| \right)
= \frac 1{\Gamma \left( \frac{s+1}2 \right)} \int_0^\infty d\tau\, \tau^{ \frac{s-1}2}
\Tr \left(  H e^{-\tau H^2} \right) .\label{eta1}
\end{equation}
Note, that zero modes are excluded from the summation in (\ref{zeta0H}). However, these modes can be included in the trace in the last expression in (\ref{eta1}) since the operator $H e^{-\tau H^2}$ vanishes on the null space of $H$.

We shall need a localized version of the formula (\ref{eta1}). To this end we introduce a smooth function $\rho$ which plays the role of a chemical potential and define
\begin{equation}
\eta (s,H;\rho)= \Tr \left( \rho \cdot (H^2)^{-s/2} H/|H| \right) .\label{etarho}
\end{equation}
Let us define a shifted Hamiltonian
\begin{equation}
H_\rho =H+\varepsilon \, \rho. \label{Hrho}
\end{equation}
so that the localized $\eta(s,H;\rho)$ function is obtained by differentiating $\eta(s,H_\rho)$ with respect to $\varepsilon$,
\begin{equation}
\eta (s,H;\rho)=-\frac 1{2\Gamma \left( \frac{s+1}2 \right)}
\int_0^\infty d\tau\, \tau^{ \frac{s-3}2} \left. \frac {\mathrm{d}}{\mathrm{d}\varepsilon} \right|_{\varepsilon=0} \, \Tr \left(  e^{-\tau H_{\rho}^2} \right). \label{eta3}
\end{equation}
Smeared and unsmeared $\eta$ functions are related through the equations
\begin{equation}
\eta (0,H;\rho)=\int d^dx \,\, \rho(x) \, \eta (0,H,x),\qquad \eta (0,H)=\int d^dx \,\, \eta(0,H;x).\label{eta5}
\end{equation}
Let us assume that there is a natural mass gap parameter $M^2$ and define
\begin{equation}
\widetilde{H}^2_\rho := H^2_\rho -M^2. \label{tHrho}
\end{equation}
For any Laplace type operator $L$ and any smooth matrix-valued function $Q$ there is an asymptotic expansion of the corresponding heat trace
\begin{equation}
\Tr \left(Q e^{-\tau L } \right)= \sum_k \tau^{\frac {k-d}2 } a_k\bigl(L;Q).
\label{hkexp}
\end{equation}
For $Q\equiv 1$ we shall abbreviate $a_k(L;1):=a_k(L)$.
If there are no boundaries, only even-numbered heat kernel coefficients $a_k(L)$ with $k=2p$ are non-zero. Clearly,
the integral in (\ref{eta3}) converges at $s=0$ if
\begin{equation}
\left. \frac {\mathrm{d}}{\mathrm{d}\varepsilon} \right|_{\varepsilon=0} \, a_k\bigl(\tilde H^2_\rho\bigr)=0\quad \mbox{for} \quad k\leq d+1 \,.\label{cond11}
\end{equation}
Let us assume that this condition is satisfied.
Then, we can substitute (\ref{hkexp}) in (\ref{eta3}), put there $s=0$ and integrate over $t$ to obtain
\begin{equation}
\eta (0,H;\rho)= -\frac 1{2\sqrt{\pi}} \sum_k \Gamma\left( \frac{k-1-d}2 \right)
|M|^{1+d-k} \left. \frac {\mathrm{d}}{\mathrm{d}\varepsilon} \right|_{\varepsilon=0} \, a_k\bigl(\tilde H^2_\rho\bigr) .\label{eta4}
\end{equation}

General operator of Laplace type can be written as
\begin{equation}
L=-(\nabla^2 +E), \qquad \nabla_\mu =\partial_\mu + \omega_\mu \label{Lap}
\end{equation}
with a suitable choice of the connection one-form $\omega$ and a matrix-valued potential $E$. In a flat space, the heat kernel coefficients are integrals of polynomials constructed from $E$, $\Omega_{\mu\nu}=[\nabla_\mu,\nabla_\nu]$ and their covariant derivatives. The general form of all heat kernel coefficients is not known. However, on solitonic backgrounds just a few structures containing a limited number of derivatives contribute to the spectral asymmetry. Such terms in $a_k(L)$ can be found for all $k$. For example, if one is interested in terms containing only $E$ and no derivatives the expression for heat kernel coefficients looks particularly simple,
\begin{equation}
a_{2p}(L;Q)=\frac 1{(4\pi)^{d/2}p!} \int d^dx \, \mathrm{tr}\, \bigl( Q E^p \bigr) .\label{a2p}
\end{equation}

Let us consider a smooth localized variation $\delta H$ of the Hamiltonian $H$. Suppose that $\delta H$ is zero order, i.e. it does not contain derivatives. Then, one can show \cite{GilkeyBook} that
\begin{equation}
\delta \eta(0,H)= -\frac 2{\sqrt{\pi}} a_{d-1}(H^2,\delta H)\,.\label{varEta}
\end{equation}
Strictly speaking, this formula is valid only if under the variation $\delta H$ eigenvalues of the Hamiltonian do not cross the zero value. In physics' literature Eq.\ (\ref{varEta}) is usually used without any restrictions (and without running into troubles because of this), see e.g. \cite{AlvarezGaume:1984nf}. Here, we assume that the background is "generic", i.e. the number of zero modes is stable under infinitesimal localized variations.

The main novelty of the approach suggested above is the use of a localizing function $\rho$. This enables us to utilize standard expressions for the heat kernel coefficients that were derived under an assumption that all integrations by parts were possible. Some other benefits of introducing $\rho$ will be discussed below, see Sec.\ \ref{sec:eta}.

\section{A non-abelian extension of the Goldstone-Wilczek model}\label{GW}
Let us consider a model in $1+1$ dimensions described by the action
\begin{equation}
S=\int dt\, dx \,\, \bar\psi \slashed{D} \psi,\qquad
\slashed{D}=\ii \gamma^\mu\partial_\mu -\varphi_1 -\ii \gamma_* \varphi_2\,,\label{GW1}
\end{equation}
where $\gamma_*=\gamma^0\gamma^1$ is the chirality matrix in $1+1$ dimensions, while $\varphi_1$ and $\varphi_2$ are scalar fields that are hermitian matrices in the "flavor" space. In the abelian one-flavor case the action (\ref{GW1}) is the model considered by Goldstone and Wilczek \cite{Goldstone:1981kk}.

The Hamiltonian for this problem reads
\begin{equation}
H=-\ii \gamma_* \partial_x +\varphi_1 \gamma^0 +\ii \varphi_2 \gamma^1\,. \label{GWH}
\end{equation}
This yields the following relation for shifted Hamiltonian (\ref{Hrho})
\begin{equation}
H_\rho^2 =-\partial_x^2+G^2 +\ii \gamma^1\varphi_1'-\gamma^0\varphi_2' +\varepsilon \bigl( 2\rho G -\ii \gamma_* (2\rho \partial_1 +\rho') \bigr),\label{GW2}
\end{equation}
where prime denotes derivative with respect to $x$ and
\begin{equation}
G= \varphi_1\gamma^0+\ii \varphi_2\,. \label{GWG}
\end{equation}
We omitted the terms with $\varepsilon^2$ since they do not contribute to the spectral asymmetry.

Take infinitesimal localized variations $\delta\varphi_1$ and $\delta\varphi_2$ so that $\delta H= \gamma^0\delta\varphi_1+\ii \gamma^1\delta \varphi_2$. From (\ref{varEta}) one computes the corresponding variation of spectral asymmetry,
\begin{equation}
\delta\eta(0,H)=-\frac 2{\sqrt{\pi}} a_0(H^2, \delta H) = -\frac 1\pi \int dx\, \mathrm{tr} \bigl(\gamma^0\delta\varphi_1+\ii \gamma^1 \delta \varphi_2\bigr)=0, \label{GWveta}
\end{equation}
due to the trace of gamma matrices. We conclude that $\eta(0,H)$ is a topological invariant. I.e., it depends on the asymptotic values of $\varphi_1$ and $\varphi_2$ only.

Let us introduce a mass gap parameter $M^2$ whose value we do no specify now. The operator $\tilde H_\rho^2$ (see eq.\ (\ref{tHrho})) can be brought to the canonical form (\ref{Lap}) with
\begin{equation}
E=(M^2-G^2) -\ii\gamma^1\varphi_1'+\gamma^0\varphi_2'-2\rho \varepsilon G,\qquad \omega_1=\ii \gamma_* \varepsilon \rho
\label{GW8}
\end{equation}
We consider only solitonic backgrounds, meaning that $\varphi_1$ and $\varphi_2$ go fast to their asymptotic values as $x\to\pm\infty$. Thus $\eta(0,H)$ depends on these asymptotic values and not on the derivatives of $\varphi_{1,2}$. Consequently, in the expansion (\ref{eta4}) we have to keep only the terms that contain at most a single first derivative of $\varphi_1$ or $\varphi_2$. This excludes the terms in the heat kernel expansion containing derivatives of $E$. Consequently, only the terms listed in (\ref{a2p}) contribute. Besides, in the powers of $E$ one should select the terms that contain $\rho$ only once. Taken together, all relevant contributions to $a_{2(l+1)}$ read
\begin{equation}
\frac 1{\sqrt{4\pi} l!} \mathrm{tr}\, \bigl[ (M^2-G^2)^l(-2\varepsilon \rho G )(-\ii \gamma^1 \varphi'_1 +\gamma^0\varphi'_2) \bigr] \label{GW9}
\end{equation}
Note, that
\begin{equation}
G^2=\varphi_1^2+\varphi_2^2+\ii \gamma_* [\varphi_1,\varphi_2].\label{GW10}
\end{equation}
After taking the trace over $\gamma$-matrices in (\ref{GW9}) we obtain
\begin{equation}
\frac{2\rho\varepsilon}{\sqrt{4\pi}l!} \mathrm{tr}\, \bigl[ (z_+^l+z_-^l) (\varphi_2\varphi'_1-\varphi_1\varphi'_2)
+\ii (z_+^l-z_-^l) (\varphi_1\varphi'_1+\varphi_2\varphi'_2) \bigr]\label{GW11}
\end{equation}
with
\begin{equation*}
z_\pm =M^2-y_\pm,\qquad y_\pm = \varphi_1^2 + \varphi_2^2\pm \ii [\varphi_1,\varphi_2].
\end{equation*}

As a next step, one has to substitute the expression (\ref{GW11}) for $a_{2(l+1)}$ in (\ref{eta4}) and compute the sum with the help of
\begin{equation}
\sum_{l=0}^\infty \frac {z_{\pm}^l}{M^{2l} }= \left( 1- \frac{z_\pm}{M^2} \right)^{-1} .\label{GWsum}
\end{equation}
This sum is convergent if $\| z_\pm \| = \| M^2 - y_\pm\| < M^2$. The matrices
\begin{equation}
y_\pm = (\varphi_1 \mp \ii \varphi_2)(\varphi_1 \mp \varphi_2)^\dag \label{GWypm}
\end{equation}
are non-negative. Thus, the problem with convergence appears only if $\varphi_1$ and $\varphi_2$ vanish simultaneously in some direction of the flavor space either in asymptotics or at some points of the bulk (meaning just a breakdown of the derivative expansion). If the problem appears at the asymptotics, one cannot do much about it. We assume that this does not happen. However, since $\eta (0,H)$ is topological, we may use the freedom of smooth variations of background fields to shift the roots of $\varphi_1$ away from the roots of $\varphi_2$ in the bulk thus making the sum (\ref{GWsum}) convergent everywhere. Thus,
\begin{equation}
\eta(0,H)=-\frac 1{2\pi} \int dx\, \mathrm{tr}\, \bigl[ (y_+^{-1}+y_-^{-1}) (\varphi_2\varphi'_1-\varphi_1\varphi'_2)
+\ii (y_+^{-1}-y_-^{-1}) (\varphi_1\varphi'_1+\varphi_2\varphi'_2) \bigr]. \label{GWeta}
\end{equation}
As expected, this expression can be converted to a boundary term
\begin{equation}
\eta(0,H)= -\frac{\ii}{2\pi} \mathrm{tr}\, \left. \bigl[ \ln (\varphi_1 +\ii \varphi_2) - \ln(\varphi_1-\ii\varphi_2) \bigr] \right|_{-\infty}^{+\infty} \label{GWb}
\end{equation}
This result is new.

If $\varphi_1$ and $\varphi_2$ are one-component fields rather than matrices, the expression (\ref{GWb}) can be simplified,
\begin{equation}
\eta(0,H)=\frac 1\pi \left. \left(\mathrm{arctg}\, \frac {\varphi_2}{\varphi_1} \right)\right|_{-\infty}^{+\infty}, \label{GWGW}
\end{equation}
which is the Goldstone and Wilczek result \cite{Goldstone:1981kk}.

We see that the method proposed works well and relatively easily. However, there are some points that have to be clarified or improved.
\begin{enumerate}
\item The heat kernel expansion reflects the behavior of heat trace at small $\tau$. To put the integration over $\tau$ on firm grounds one needs to understand the behavior at large $\tau$ as well.
\item It is hard to control the rate of convergence of the expansion (\ref{GWeta}), especially since the individual terms in (\ref{GWeta}) do not give total derivatives after the variation with respect to $\rho$.
\item In the example above we had to sum up the simplest contributions $E^k$ to the heat kernel coefficients. In more complicated cases, especially in higher dimensions, more complicated invariants may become relevant. Thus, it is useful to have a method to find all higher heat kernel coefficients for some types of the operators.
\end{enumerate}
These points will be addressed below.

\section{The kink fermion number through (generalized) heat kernel expansion}\label{sec:kink}
\subsection{The spectrum of fluctuations}\label{sec:spec}
In this Section, we demonstrate how the heat kernel expansion can be improved by taking into account some global information on the background fields. Below, we describe the properties of Dirac spectrum on kink backgrounds (cf \cite{Niemi:1984vz}). We take
\begin{equation}
H=H_0 + \mu \sigma_3 = \left( \begin{array}{cc} \mu & -\frac{d}{dx}+\Phi(x) \\
\frac{d}{dx}+\Phi(x) & -\mu \end{array} \right) = \left( \begin{array}{cc} \mu & D \\
D^\dagger & -\mu \end{array} \right) \label{DiracOperator}
\end{equation}
where $\mu$ is the mass, and $\Phi(x)$ refers to a yet unspecified topological kink with a smooth behavior. One can think of $\Phi=\nu \mathrm{tanh}(x)$, though we shall not restrict ourselves to this simplest example. The Hamiltonian (\ref{DiracOperator}) is a particular case of (\ref{GWH}). The eigenspinors of  (\ref{DiracOperator}) satisfy
the Dirac equation:
\begin{equation}
\left( \begin{array}{cc} \mu & D \\ D^\dagger & -\mu \end{array} \right) \left( \begin{array}{c} u_1(x) \\ u_2(x) \end{array} \right) =
E \left( \begin{array}{c} u_1(x) \\ u_2(x) \end{array} \right) \, \, \, \,  \equiv \quad \left\{\begin{array}{c} D^\dagger u_1(x) = (E+\mu)u_2(x) \\  Du_2(x)=(E-\mu)u_1(x) \end{array}\right.
 \label{DiracEquation}
\end{equation}
Finding the solution of the ODE system (\ref{DiracEquation}) if $\vert E\vert\neq \vert \mu\vert$  is equivalent to solving one spectral Schr$\ddot{\rm o}$dinger problem in two ways:
\begin{equation}
\left\{\begin{array}{c} DD^\dagger u_1(x) = (E^2-\mu^2) u_1(x)\\ u_2(x)=\frac{1}{E+\mu}D^\dagger u_1(x) \end{array}\right. \hspace{0.9cm},\hspace{0.9cm} \left\{\begin{array}{c} D^\dagger D u_2(x) =(E^2-\mu^2) u_2(x)\\ u_1(x)=\frac{1}{E-\mu}Du_2(x)\end{array}\right.\label{dirsol}
\end{equation}
where the  Schr$\ddot{\rm o}$dinger operators are defined in terms of the topological kink configuration:
\begin{eqnarray}
DD^\dagger  &=& -\frac{d^2}{dx^2} - \frac{d\Phi}{dx} + \Phi^2(x) = -\frac{d^2}{dx^2} + v^2 + V_1(x) \label{dda} \\
D^\dagger D &=& -\frac{d^2}{dx^2} + \frac{d\Phi}{dx} + \Phi^2(x) = -\frac{d^2}{dx^2} + v^2 + V_2(x) \label{dad}
\end{eqnarray}
In (\ref{dda})-(\ref{dad}) we have taken into account that kink configurations have very definite asymptotic behavior:
\[
\lim_{x\rightarrow \pm \infty} \Phi^2(x)=v^2 \quad , \quad \lim_{x\to\pm\infty}\frac{d\Phi}{dx}=0
\]
where $v$ is a constant. Thus, $V_1(x)=\Phi^2(x)-v^2-\frac{d\Phi}{dx}$ and $V_2(x)=\Phi^2(x)-v^2+\frac{d\Phi}{dx}$ also tend to zero at $x=\pm\infty$:
\[
\lim_{x\rightarrow \mp \infty} V_1(x) = \lim_{x\rightarrow \mp \infty} V_2(x) =0
\]
We summarize the spectrum of the Dirac operator:

\begin{enumerate}
	\item If $E=\mu$ then $D^\dagger u_1(x)= 2\mu u_2(x)$ and $Du_2(x)=0$. Henceforth $DD^\dagger u_1(x)=0$, which means that $u_1$ is a zero mode of the operator $DD^\dagger$ if it is normalizable.
	
	\item If $E=-\mu$ then $D^\dagger u_1(x)=0$ and $D u_2(x)=-2\mu u_1(x)$. Henceforth $D^\dagger D u_2(x)=0$, which means that $u_2(x)$ in this case is a zero mode of the operator $D^\dagger D$ if normalizable.
We stress that only the spinor annihilated by  either $D^\dagger$ or by $D$ is normalizable. Therefore, generically, for one-scalar field theory there will only be one normalizable zero mode of the Dirac operator governing
the spinorial fluctuations around one kink.
	
	\item If $E\neq \pm \mu$ the first component of the spinor $u_1(x)$ is obtained by solving the spectral problem
	\begin{equation}
	DD^\dagger u_1(x) =(E^2 - \mu^2) u_1 = \omega_1^2 u_1(x) \label{SpectralProblem1}
	\end{equation}
	whereas the second component $u_2$ is fixed in (\ref{dirsol})-(left).
    Clearly this solution for $u_2$ is an eigenfunction of the spectral problem
	\begin{equation}
	D^\dagger D u_2(x) =(E^2 - \mu^2) u_2(x) = \omega_2^2 u_2(x) \label{SpectralProblem2}
	\end{equation}
	because the intertwining of the $DD^\dagger$ and $D^\dagger D$ operators which , except the zero mode, are isospectral:
	\[
	D^\dagger D u_2= \frac{1}{E+\mu} D^\dagger D D^\dagger u_1 = \frac{1}{E+\mu} D^\dagger (E^2 - \mu^2) u_1 = (E-\mu) D^\dagger u_1 =(E^2-\mu^2) u_2
	\]
	Denoting as $\widetilde{u}_1$ a normalized eigenfunction of the spectral problem (\ref{SpectralProblem1}) $DD^\dagger u_1(x) = \omega^2 u_1(x)$ one easily checks that the spinor
	\begin{equation}
	u(x)=\left( \begin{array}{c}  \sqrt{\frac{E+\mu}{2E}} \, \widetilde{u}_1(x) \\ \frac{{\rm sign}(E)}{\sqrt{2E(E+\mu)}} D^\dagger \widetilde{u}_1(x) \end{array} \right) \label{spinor01}
	\end{equation}
	is a normalized eigenfunction of the Dirac Hamiltonian, since via a shrewd partial integration one finds that: $\int dx \, u^\dagger(x) u(x) = \int dx \, \widetilde{u}_1^*(x) \widetilde{u}_1(x)$.
	
	The function
	\begin{equation}
	\widetilde{u}_2 (x)= \sqrt{\frac{E+\mu}{E-\mu}} u_2(x)= \frac{1}{\sqrt{E^2-\mu^2}} D^\dagger \widetilde{u}_1(x) \label{u2normalized}
	\end{equation}
	is obviously an eigenfunction of the spectral problem (\ref{SpectralProblem2}) $D^\dagger D u_2(x)=(E^2-\mu^2)u_2(x)$ because it is proportional to $u_2$ and, after another partial integration, one checks that its norm is one: $\int dx \, \widetilde{u}_2^* \widetilde{u}_2 = \int dx \, \widetilde{u}_1^* \widetilde{u}_1$.

We thus may characterize the eigenspinors of the Dirac Hamiltonian as
	\begin{equation}
	u_E(x)= \left(\begin{array}{c} \sqrt{\frac{E+\mu}{2E}} \, \widetilde{u}_1(x,E) \\ \sqrt{\frac{E-\mu}{2E}} \, {\rm Sign}(E) \,\widetilde{u}_2(x,E) \end{array} \right) \label{spinor02}
	\end{equation}
instead of (\ref{spinor01}). To distinguish among the different eigenspinors we remark that $\widetilde{u}_1(x,E)$ and $\widetilde{u}_2(x,E)$ are, respectively, non-null normalized eigenfunctions of $DD^\dagger$ and $D^\dagger D$. They both are characterized by either a real vector number $k\in{\mathbb R}$ if one search for scattering states
and/or purely imaginary momenta, $k=i\kappa_j$ with $\kappa_j>0$ and $j=1,2, \cdots , N_B$, if $N_B$ bound states exist. The corresponding energies are: $E^{(\pm)} = \pm\sqrt{k^2+v^2+\mu^2}$, and $ E_j^{(\pm)}=\pm\sqrt{v^2+\mu^2-\kappa_j^2}$. Besides there is a zero mode $\kappa_0=v$ of either $DD^\dagger$ or $D^\dagger D$ which contributes to the spectral asymmetry selecting either $E_\mu=\mu$ or $E_{-\mu}=-\mu$.
We denote by $B_1=\{ \widetilde{u}_{1k} \}_{k\in I}$ the complete orthogonal set in $L^2({\mathbb R})$ formed by the
eigenfunctions of $DD^\dagger$. $I$ is a set of indices that symbolically label the scattering states, the bound states, and, eventually, one zero mode. Also we dispose of $B_2=\{u_{2k}\}_{k\in I}$ as a second complete orthogonal basis in $L^2({\mathbb R})$ formed by the eigenfunctions of the operator $D^\dagger D$, except for the possible a priori presence of a zero mode.
\end{enumerate}

Most of the properties established here are also valid in higher dimension \cite{Chamon:2007hx}.

\subsection{Niemi-Semenoff formula and the spectral heat trace}\label{sec:NS}
The fermionic quantum field in the Schr$\ddot{\rm o}$dinger picture is expanded in terms of the Dirac eigenspinors:
\begin{equation}
\widehat{\psi}(x)=\int \, [dk] \, \left(\hat{b}_+(k)u_{E_+}(x)+\hat{b}^\dagger_-(k)u_{E_-}(x)\right)\quad .
\end{equation}
Here $[dk]$ denotes an integration measure on the spectrum of the Dirac operator. The fermionic creation and annihilation operators of particle and antiparticles $\hat{b}_\pm(k)$, $\hat{b}^\dagger_\pm(k)$
satisfies the anticommutation relations:
\begin{equation}
\left\{\hat{b}_+(k),\hat{b}_+^\dagger(q)\right\}=\delta(k-q)=\left\{\hat{b}_-(k),\hat{b}_-^\dagger(q)\right\}\, \, .
\end{equation}
All other anticommutators between these operators are zero. The one-particle states may be therefore only occupied by one Fermion or unoccupied and one state is distinguished as the ground state: $\big\vert 0 \big\rangle$, characterized
as the vacuum state where all the one-particle states are unoccupied, $\hat{b}_+(k)\big\vert 0\big\rangle=\hat{b}_-(k)\big\vert 0\big\rangle=0$, $\forall k$. From the expectation value at the vacuum state
of the normal ordered Fermi density, one has
\begin{eqnarray}
\big\langle 0 \big\vert : \hat{\rho}(x) : \big\vert 0 \big\rangle &=&\frac{1}{2}\big\langle 0\big\vert \left(\hat{\psi}^\dagger(x)\hat{\psi}(x)-\hat{\psi}(x)\hat{\psi}^\dagger(x)\right)\big\vert 0\big\rangle \nonumber\\
&=& \frac{1}{2}\int \, [dk] \, \left(\hat{b}_+(k)\hat{b}_+^\dagger(k)u^\dagger_{E_+}(x)u_{E_+}(x)-\hat{b}_-(k)\hat{b}_-^\dagger(k)u^\dagger_{E_-}(x)u_{E_-}(x)\right) \, \, .
\end{eqnarray}
This calculation has been performed using the anticommutation relations. Further use of the anticommutation relations allows us to obtain
the fermionic number of the kink ground state:
\begin{eqnarray}
N&=& \int_{-\infty}^\infty \, dx \, \big\langle 0 \big\vert : \hat{\rho}(x) : \big\vert 0 \big\rangle \nonumber \\ &=& -\frac{1}{2} \Big[ \int_{-\infty}^\infty dx \int \, [dk] u_E(x)^\dagger u_E(x) -  \int_{-\infty}^\infty dx \int \, [dk] u_{-E}(x)^\dagger u_{-E}(x) \Big]  \label{densF}
\end{eqnarray}
Formula (\ref{spinor02}) permits to rewrite (\ref{densF}) in the form:
\begin{eqnarray}
N&=&-\frac{1}{2} \Big[ \int_{-\infty}^\infty dx \int \, [dk] \frac{\mu}{E} \widetilde{u}_1(x)^* \widetilde{u}_1(x) +  \int_{-\infty}^\infty dx \int \, [dk] \Big( -\frac{\mu}{E} \Big) \widetilde{u}_{2}(x)^* \widetilde{u}_{2}(x) \Big]	= \nonumber\\
&=& -\frac{1}{2} \Big[ \int \, [dk] \frac{\mu}{\sqrt{\omega_{1k}^2 + \mu^2}} \int_{-\infty}^\infty \widetilde{u}_1^* \widetilde{u}_1 dx -  \int \, [dk] \frac{\mu}{\sqrt{\omega_{2k}^2 + \mu^2}} \int_{-\infty}^\infty \widetilde{u}_2^* \widetilde{u}_2 dx \Big] = \nonumber\\
&=& -\frac{1}{2} \Big[  \int \, [dk] \, \frac{\mu}{\sqrt{\omega_{1k}^2+\mu^2}} - \int \, [dk] \frac{m}{\sqrt{\omega_{2k}^2+ \mu^2}} \Big] \label{kinkfermi}
\end{eqnarray}
Each \lq\lq integral\rq\rq{} in this formula for $N$ is ultraviolet divergent, $v^2$ and $\mu^2$, however, prevent infrared divergences. To control the ultraviolet divergences we shall use spectral zeta function/heat trace methods associated with the operators $DD^\dagger$ and $D^\dagger D$.

We put at work this strategy in a more general context. Translate the spectral zeta function
corresponding to the second order differential operator{\footnote{ Note that this differential operator is a 1D
member of the family of Laplace type operators (\ref{Lap}).}}
\[
{\cal K}= -\frac{d^2}{dx^2} + v^2 + V(x)
\]
to its spectral heat trace via the Mellin's transform:
\begin{eqnarray}
&&{\rm Tr} \left[ \frac{1}{({\cal K}+\mu^2)^s} - \frac{1}{({\cal K}_0+\mu^2)^s} \right] = \frac{1}{\Gamma[s]} \int_0^\infty \bbeta^{s-1} {\rm Tr}\left[ e^{-\bbeta ({\cal K}+\mu^2)} -e^{-\bbeta ({\cal K}_0+\mu^2)}\right]
\nonumber\\
&&\qquad\qquad\qquad \equiv \frac{1}{\Gamma[s]} \int_0^\infty \bbeta^{s-1} e^{-\bbeta \mu^2} h_{\cal K}(\bbeta) d\bbeta.
\end{eqnarray}
To tame ultraviolet divergences we subtracted the contribution from ${\cal K}_0$, which is obtained from ${\cal K}$ by putting $V=0$. The operator ${\cal K}_0$ is the same for $DD^\dagger$ and $D^\dagger D$, so that the correction term is canceled in $N$. We denote $h_{\cal K}(\bbeta)= {\rm Tr} \left[ e^{-\bbeta \,{\cal K}} -e^{-\bbeta \,{\cal K}_0}\right] $. The kink fermion number as given in formula (\ref{kinkfermi}) may be written in terms of the spectral heat functions of the paired Schr$\ddot{\rm o}$dinger operators as follows:
\[
N= -\frac{\mu}{2} \Big[ \lim_{s\rightarrow \frac{1}{2}} \frac{1}{\Gamma[s]} \int_0^\infty \bbeta^{s-1} e^{-\bbeta \mu^2} h_{DD^\dagger}(\bbeta) d\bbeta - \lim_{s\rightarrow \frac{1}{2}} \int_0^\infty \bbeta^{s-1} e^{-\bbeta \mu^2} h_{D^\dagger D}(\bbeta) d\bbeta \Big]
\]
At this point we use the asymptotic expansion of the kink spectral trace, see \cite{Alonso:2012},
\begin{equation}
h_{\cal K}(\bbeta) = \sum_{n=1}^\infty c_n({\cal K}) e^{-\bbeta v^2} \frac{1}{\sqrt{4\pi}} \bbeta^{n-\frac{1}{2}} + N_{zm}^{\cal K} \, {\rm erf} (v\sqrt{\bbeta}) \label{htae} \,.
\end{equation}
Here, $N_{\rm zm}^{\cal K}$ is the number zero modes in the spectrum of ${\cal K}$. The error function in equation (\ref{htae}) is introduced to guarantee the right behavior of the heat trace
at low temperature, $\bbeta\to \infty$, when zero modes are present, whereas the high-temperature asymptotics is preserved. Any regular function with the same limits as the error function at $\bbeta\to \infty$ and $\bbeta\to 0$ could do the job but our choice optimizes the solution of the recurrence relations, see \cite{Alonso:2012}. In addition, this procedure implies a natural relation between the Seeley coefficients of two intertwined Laplace type operators Darboux factorizable in terms of one ladder operator and its adjoint. The coefficients $c_n({\cal K})$ are obtained by solving the recurrence relations arising in the power series expansion solution of the ${\cal K}$-heat equation in the so-called modified Gilkey-de Witt approach, see \cite{Alonso:2012}.
The coefficients $c_n$ play a role similar to that of $a_{2n}$ is Sec.\ \ref{sec:hk}, apart from normalization and modification of the expansion.

Application to the kink fermion number formula gives
\begin{equation}
N= -\frac{\mu}{2} \Big[ \lim_{s\rightarrow \frac{1}{2}} \frac{1}{\Gamma(s)} \Big\{ \int_0^\infty \bbeta^{s-1} e^{-\bbeta \mu^2} \Big[h_{DD^\dagger}(\bbeta) - h_{D^\dagger D}(\bbeta) \Big] d\bbeta  \Big\} \Big] \nonumber \, ,
\end{equation}
or, after performing the asymptotic expansion and subsequently the Mellin transform we obtain
\begin{eqnarray*}
N&=& -\frac{\mu}{2} \Big[ \lim_{s\rightarrow \frac{1}{2}} \frac{1}{\Gamma[s]}\Big\{ \sum_{n=1}^\infty \frac{c_n(DD^\dagger)}{\sqrt{4\pi}} (\mu^2+ v^2)^{\frac{1}{2}-n-s} \, \Gamma[n+s-{\textstyle \frac{1}{2}}] +  \\
&& \hspace{0.8cm} + N_{zm}^{DD^\dagger} \, \frac{2v}{\sqrt{\pi}} \, (\mu^2)^{-\frac{1}{2}-s} \Gamma[s+{\textstyle \frac{1}{2}}] \, {}_2F_1 [{\textstyle \frac{1}{2}},{\textstyle \frac{1}{2}}+ s,{\textstyle \frac{3}{2}} ,-{\textstyle\frac{v^2}{\mu^2}}] - \\
&&\hspace{0.8cm} - \sum_{n=1}^\infty \frac{c_n(D^\dagger D)}{\sqrt{4\pi}} \, (\mu^2+ v^2)^{\frac{1}{2}-n-s} \, \Gamma[n+s-{\textstyle \frac{1}{2}}] +  \\
&& \hspace{0.8cm}- N_{zm}^{D^\dagger D} \, \frac{2v}{\sqrt{\pi}} \, (\mu^2)^{-\frac{1}{2}-s} \, \Gamma[s+{\textstyle \frac{1}{2}}] \, {}_2F_1 [{\textstyle \frac{1}{2}},{\textstyle \frac{1}{2}}+s,{\textstyle \frac{3}{2}},-{\textstyle \frac{v^2}{\mu^2}}]  \Big\} \Big]
\end{eqnarray*}
Finally, taking the limit in the previous expression, one obtains the formula
\[
N=-\frac{\mu}{4\pi} \sum_{n=1}^\infty \left( c_n(DD^\dagger)- c_n(D^\dagger D) \right)\, \frac{(n-1)!}{(\mu^2+v^2)^n} - \frac{1}{\pi} \arctan \Big(\frac{v}{\mu} \Big) \Big( N_{zm}^{DD^\dagger} - N_{zm}^{D^\dagger D} \Big) \, ,
\]
which is closely related to the Niemi-Semenoff formula.

In the case of kink type operators where $DD^\dagger$ and $D^\dagger D$ are intertwined operators it may be shown that $c_n(DD^\dagger)=c_n(D^\dagger D)$, $\forall n$ when the modified Gilkey-de Witt approach is applied. We recall that this scheme is adapted to the presence of zero modes in the spectrum of a operator. We shall prove this fact for a
pair of intertwined operators having an spectral structure identical to the spectral structures of $DD^\dagger$ and $D^\dagger D$ in the kink background. For the sake of simplicity it will be assumed that the operator $DD^\dagger$ encompasses a zero mode whereas $D^\dagger D$ has a positive spectrum. Otherwise, the following result remains true swapping the r$\hat{\rm o}$le of the first-order operators $D$ and $D^\dagger$.

\vspace{0.2cm}

\noindent \textsc{Lemma:} Let $D=-\frac{d}{dx}+\Phi(x)$ be a first-order differential operator such that $\lim_{x\rightarrow \pm \infty} \Phi(x)=\pm v$ and $\psi_0(x)=e^{-\int_x \Phi(x) dx}$ is a square-integrable function annihilated by $D^\dagger$. The Seeley coefficients of the operators $D D^\dagger$ and $D^\dagger D$ of the spectral $DD^\dagger$-and $D^\dagger D$-heat functions in the improved Gilkey-de Witt  expansion coincide: $c_n(D^\dagger D) = c_n (D D^\dagger)$ $\forall n\in \mathbb{N}$.

\vspace{0.2cm}

\noindent \textsc{Proof:} We shall use the notation given by (\ref{dda}) and (\ref{dad}). Let $\{f_k(x)\}$ and $\{g_k(x)\}$ be complete bases of orthonormal eigenfunctions of the operators $D D^\dagger$ and $D^\dagger D$, respectively. We summarize the spectra of these operators without taking into account yet that they are intertwined and, thus, except for the zero modes, they are isospectral
\begin{eqnarray*}
{\rm Spec}(D D^\dagger) &=&\{0\} \cup \{\omega_{1i}^2\}_{i=1}^{N_1} \cup \{k^2+v^2\}_{k\in \mathbb{R}} \\
{\rm Spec}(D^\dagger D) &=&  \{\omega_{2j}^2 \}_{j=1}^{N_2} \cup \{k^2+v^2\}_{k\in \mathbb{R}}
\end{eqnarray*}
In terms of the associated eigenfunctions the heat kernels of these operators are expressed as the (formal) series:
\begin{eqnarray*}
K_{D D^\dagger}(x,y;\bbeta) &=& f_0^*(y)f_0(x)+ \sum_{i=1}^{N_1} f_i^*(y) f_i(x) e^{-\bbeta \omega_{1i}^2} + \int_{-\infty}^\infty \, dk \rho_{DD^\dagger}(k) f_k^*(y) f_k(x) e^{-\bbeta \omega_{1k}^2(k)} \\
K_{D^\dagger D} (x,y;\bbeta)&=&  \sum_{i=1}^{N_2} g_i^*(y) g_i(x) e^{-\bbeta \omega_{2i}^2} + \int_{-\infty}^\infty dk \, \rho_{D^\dagger D}(k) g_k^*(y) g_k(x) e^{-\bbeta \omega_{2k}^2(k)}
\end{eqnarray*}
where $f_0(x)$ is the zero mode of the operator $DD^\dagger$.

The action of the \lq\lq ladder\rq\rq{} operators $D^\dagger$ and $D$ respectively on the spectral equations for $D D^\dagger$ and $D^\dagger D$ easily show isospectrality of the positive spectra: $N_1=N_2$ and $\omega_{1i}^2 = \omega_{2i}^2 =\omega_i^2$ with $i\geq 1$. In fact $g_k(x)=\frac{1}{\omega(k)} D^\dagger f_k(x)$ are normalized eigenfunctions of the spectral the operator $D^\dagger D$. A caveat: since the mapping is via a differential operator the spectral densities of the scattering states are different. The heat equation kernel of the $D^\dagger D$ operators can be now written in terms of the eigenfunctions of the $DD^\dagger$ operator as
\begin{eqnarray*}
K_{D^\dagger D}(x,y;\bbeta) &=& \sum_{i=1}^{N} (D^\dagger f_i)^*(y) (D^\dagger f_i(x)) \frac{1}{\omega_i^2} e^{-\bbeta \omega_{i}^2} + \\ & +&  \int_{-\infty}^\infty dk\,\rho_{D^\dagger D}(k) (D^\dagger f_k)^*(y) (D^\dagger f_k(x)) \frac{1}{\omega(k)^2} e^{-\bbeta \omega_{k}^2(k)}
\end{eqnarray*}
The trace of this kernel $h_{D^\dagger D}(\bbeta) = \int dx \, K_{D^\dagger D}(x,x;\bbeta)$ may be written thus
\begin{eqnarray*}
	&& h_{D^\dagger D}(\bbeta) =  \sum_{i=1}^{N'} \int dx f_i^*(x) D D^\dagger f_i(x) \frac{1}{\omega_i^2} e^{-\bbeta \omega_{i}^2} + \\ & & +  \int_{-\infty}^\infty dk\, \rho_{D^\dagger D}(k) \int dx f_k^*(x) DD^\dagger f_k(x) \frac{1}{\omega(k)^2} e^{-\bbeta \omega_{k}^2(k)} =
\sum_{i= 1}^N e^{-\bbeta \omega_i^2} + \int dk\, \rho_{D^\dagger D}(k) \, e^{-\bbeta \omega^2(k)}
\end{eqnarray*}
Moreover $\rho_{DD^\dagger}(k) = \rho_{D^\dagger D}(k)
- \frac{1}{\pi} \frac{v}{k^2 + v^2}$, see \cite{Alonso:2013}, which implies
\begin{eqnarray*}
	h_{D^\dagger D} (\bbeta) & =& \sum_{i=1}^N e^{-\bbeta \omega_i^2} + \int_{-\infty}^\infty \Big[ \rho_{DD^\dagger}(k) + \frac{1}{\pi} \frac{v}{k^2+v^2}  \Big] e^{-\bbeta \omega^2(k)} = \\
	&=& h_{DD^\dagger} -1 + \frac{v}{\pi} \int_{-\infty}^\infty dk \frac{1}{k^2+v^2} e^{-\bbeta(k^2+v^2)} =  \\
	&=& h_{DD^\dagger}(\bbeta) - {\rm erf}(v\sqrt{\bbeta})
\end{eqnarray*}
that is,
\[
h_{D^\dagger D} (\bbeta)=h_{DD^\dagger}(\bbeta) - {\rm erf}(v\sqrt{\bbeta})
\]
Plugging the modified Gilkey-de Witt expansion of the heat function for these operators
\begin{eqnarray*}
h_{D^\dagger D}(\bbeta) &=& \sum_{n=1}^\infty c_n(D^\dagger D) \, \frac{1}{\sqrt{4\pi}} \, e^{-\bbeta v^2} \bbeta^{n-\frac{1}{2}} \\
h_{D D^\dagger}(\bbeta) &=& \sum_{n=1}^\infty c_n(D D^\dagger) \, \frac{1}{\sqrt{4\pi}} \, e^{-\bbeta v^2} \bbeta^{n-\frac{1}{2}} + \, {\rm erf}(v\sqrt{\bbeta})
\end{eqnarray*}
into the previous expression
\[
\sum_{n=1}^\infty c_n(D^\dagger D) \, \frac{1}{\sqrt{4\pi}} \, e^{-\bbeta v^2} \bbeta^{n-\frac{1}{2}} = \sum_{n=1}^\infty c_n(D D^\dagger) \, \frac{1}{\sqrt{4\pi}} \, e^{-\bbeta v^2} \bbeta^{n-\frac{1}{2}}
\]
leads to the result
\[
c_n(D^\dagger D) = c_n (D D^\dagger) \hspace{0.5cm}\mbox{with} \hspace{0.5cm} n\in \mathbb{N}
\]
as claimed $\Box$.

\vspace{0.2cm}

In sum, the Niemi-Semenoff formula for this type of Dirac operators hold:
\[
N=-\frac{1}{\pi}\arctan \frac{v}{\mu} \Big( N_{zm}^{DD^\dagger} - N_{zm}^{D^\dagger D} \Big)
\]
where $N_{zm}^{DD^\dagger} - N_{zm}^{D^\dagger D} = \pm 1$.

The fermion number computed at the kink ground state, with all the fermionic fluctuations around the kink being unoccupied, is an angle rather than an integer number- in fact zero- as it would be in the usual vacuum state. In the $\mu=0$ limit the Jackiw-Rebbi half-integer fermion number is recovered.

\subsection{Fermion number from the spectral eta function}\label{sec:eta}
Here we evaluate the fermion number by using the expression (\ref{eta3}) for spectral asymmetry. Let us write
\[
H_\rho^2 = -\frac{d^2}{dx^2} \mathbb{I} + v^2 + V(x) + Q(x) \frac{d}{dx} \, \, ,
\]
where $V(x)$ and $Q(x)$ are matrix-valued functions that vanish at $x\to\pm\infty$, $v^2$ is a diagonal matrix representing asymptotic values of the potential, $v^2 ={\rm diag} (v_1^2,v_2^2)$, and $\mathbb{I}$ is a unit matrix. All matrices are $2\times 2$. An extension to matrices of arbitrary size is straightforward.

Let us take an expansion
\[
{\rm Tr} \Big( e^{-\bbeta H_\rho^2} \Big) = \sum_{n=0}^\infty \sum_{i=1}^2 \, [c_n(H_\rho^2)]_{ii} \, e^{-\bbeta v_i^2}\, \frac{1}{\sqrt{4\pi}}\, \bbeta^{n-\frac{1}{2}}, \label{hkhk}
\]
which essentially coincides with the one that we used in Sec. \ref{sec:hk} after obvious changes in the notations. In the expansion (\ref{hkhk}), however, the mass gap $v^2$ need not be proportional to a unit matrix. After plugging this expansion in (\ref{eta3}), integrating over $\bbeta$ and passing to $s=0$, we obtain
\begin{equation}
N(H) = \frac{1}{8\pi} \sum_{n=0}^\infty \sum_{i=1}^2 \,[\overline{c}_n(H_\rho^2)]_{ii} \, (v_i^2)^{1-n} \, \Gamma[n-1] \label{fermionicnumber02} \, \, ,
\end{equation}
where we denoted
\[
[\overline{c}_n(H_\rho^2)]= \lim_{\rho(x)\rightarrow 1} \,\frac{d}{d\varepsilon} \Big|_{\varepsilon=0} [c_n(H_\rho^2)]\quad\, \, .
\]
To extract the Seeley coefficients $[c_n(H_\rho^2)]_{ii}$ we first solve the recurrence relations{\footnote{See Reference \cite{Guilarte:2014}, Section \S.3, to find the notational details and the developments of the main steps in this derivation.}} for the derivatives of the diagonal densities ${}^{(k)}C_n(x)=\lim_{y\to x}\frac{\partial^k}{\partial x^k}c_n(x,y)$
\begin{eqnarray*}
&& {}^{(k)}C_n(x) = \frac{1}{n+k} \Big[ {}^{(k+2)}C_{n-1}(x) - \sum_{j=0}^k {k \choose j} \frac{d^j V(x)}{d x^j}\, {}^{(k-j)}C_{n-1}(x) - [v^2 , {}^{(k)}C_{n-1}(x)] - \\
&& \hspace{1cm} - \sum_{j=0}^k {k\choose j} \frac{d^j Q(x)}{d x^j} \, {}^{(k-j+1)}C_{n-1}(x) + \frac{k}{2} \sum_{j=0}^{k-1} {k-1 \choose j} \frac{d^j Q(x)}{d x^j} \, {}^{(k-j-1)}C_{n}(x) \Big]
\end{eqnarray*}
starting with the initial conditions ${}^{(k)} C_0(x) = \delta_{0k} \, \mathbb{I}$.
The Seeley coefficients are thus obtained from the diagonal densities via integration
\[
c_n(H_\rho^2) = \int_{-\infty}^\infty dx \, {}^{(0)}C_n(x)
\]
and the fermionic number is estimated as the $n_{\rm max}\to \infty$ limit of the partial sums:
\[
N^{(n_{\rm max})}= \frac{1}{8\pi} \sum_{n=0}^{n_{\rm max}} \sum_{i=1}^N \, [\overline{c}_n(H_\rho^2)]_{ii} \, (v_i^2)^{1-n}\, \Gamma[n-1]
\]
If the $v_i$ are all equal we have
\[
N^{(n_{\rm max})}= \frac{1}{8\pi} \sum_{n=0}^{n_{\rm max}} {\rm tr} [\overline{c}_n(H_\rho^2)] \, (v^2)^{1-n}\, \Gamma[n-1]
\]
Recall that the square of the Niemi-Semenoff-Dirac Hamiltonian (\ref{DiracOperator}) is
\[
H_\rho^2= \left( \begin{array}{cc} -\frac{d^2}{dx^2} -\frac{d\Phi(x)}{dx} + \Phi^2(x) +(\mu+\varepsilon \rho(x))^2 & -2\varepsilon \rho(x) \frac{d}{dx} - \epsilon  \frac{d\rho(x)}{dx} + 2\epsilon \rho(x)\Phi(x) \\2\varepsilon \rho(x) \frac{d}{dx} + \epsilon \frac{d\rho(x)}{dx} + 2\epsilon \rho(x)\Phi(x)  & -\frac{d^2}{dx^2} +\frac{d\Phi(x)}{dx} + \Phi^2(x) +(-\mu+\varepsilon \rho(x))^2 \end{array} \right)
\]
where $\Phi(x)$ is the solitonic background which complies with the asymptotic behavior
\[
\lim_{x\rightarrow \pm \infty} \Phi(x)= \pm \nu
\]
Therefore, we have $v^2= {\rm diag}(\nu^2+\mu^2,\nu^2+\mu^2)$,
\[
V(x) = \left( \begin{array}{cc}  -\frac{d\Phi(x)}{dx} + \Phi^2(x)- \nu^2 + 2\mu \varepsilon \rho(x)+\varepsilon^2 \rho^2(x) & \epsilon  \frac{d\rho(x)}{dx} + 2\epsilon \rho(x)\Phi(x) \\  - \epsilon \frac{d\rho(x)}{dx} + 2\epsilon \rho(x)\Phi(x)  & \frac{d\Phi(x)}{dx} + \Phi^2(x) - \nu^2 - 2\mu \varepsilon \rho(x)+\varepsilon^2 \rho^2(x) \end{array} \right)
\]
and
\[
Q(x) = \left( \begin{array}{cc} 0 & 2\varepsilon \rho(x) \\ -2\varepsilon \rho(x) & 0 \end{array} \right)
\]
Plugging these expressions into the recurrence relations the solutions for the lower Seeley coefficients provided by Mathematica read:
\begin{eqnarray*}
&& {\rm tr}[\overline{c}_0(H_\rho^2)]=  0 \hspace{0.3cm},\hspace{0.3cm} {\rm tr}[\overline{c}_1(H_\rho^2)]=  0 \hspace{0.3cm},\hspace{0.3cm}  {\rm tr}[\overline{c}_2(H_\rho^2)]=  -8 \, \mu \, \nu \hspace{0.3cm},\hspace{0.3cm}  {\rm tr}[\overline{c}_3(H_\rho^2)]=  -\frac{16}{3} \,\mu \,\nu^3 , \\
&& {\rm tr}[\overline{c}_4(H_\rho^2)] =  -\frac{32}{15} \, \mu \, \nu^5 \hspace{0.2cm},\hspace{0.2cm}  {\rm tr} [\overline{c}_5(H_\rho^2)]= -\frac{64}{105} \, \mu \, \nu^7 \hspace{0.2cm},\hspace{0.2cm} {\rm tr} [\overline{c}_6(H_\rho^2)]= -\frac{128}{945} \, \mu \, \nu^9  \hspace{0.2cm},\hspace{0.2cm} \dots
\end{eqnarray*}
For example,
\begin{eqnarray*}
&& {\rm tr}[\overline{c}_2(H_\rho^2)]= \int_{-\infty}^\infty dx (-4 \mu \Phi'(x)) = -4\mu \Big[\Phi(+\infty)-\Phi(-\infty)\Big] =  -8 \, \mu \, \nu \hspace{0.2cm} , \\
&& {\rm tr}[\overline{c}_3(H_\rho^2)] =  -\frac{2}{3} \mu\int_{-\infty}^\infty dx \Big[  6(\nu^2-\Phi^2(x)) \Phi'(x) +\Phi'''(x) \Big] = \\
&& \hspace{0.8cm} = -\frac{2}{3} \mu \Big[ 6 \Big(\nu^2 \Phi(x)- \frac{1}{3} \Phi^3(x)\Big) + \Phi''(x) \Big] \Big|_{-\infty}^\infty =   -\frac{16}{3} \,\mu \,\nu^3 ,\\
&& {\rm tr}[\overline{c}_4(H_\rho^2)] = \mu \int_{-\infty}^\infty dx \Big[ -2 (\nu^2 - \Phi^2(x))^2 \Phi'(x) + \frac{2}{3} \Phi'(x)^3+ \\
&& \hspace{1.2cm} + \frac{8}{3} \Phi(x) \Phi'(x) \Phi''(x) - \frac{2}{3} \Big( \nu^2 - \Phi^2(x) \Big) \Phi'''(x)-\frac{1}{15} \Phi^{(5)}(x) \Big] =\\
&& \hspace{0.8cm} = \mu \Big(-2\nu^4 \Phi(x)+\frac{4}{3} \nu^2 \Phi^3(x) - \frac{2}{5} \Phi^5(x) \Big) \Big|_{-\infty}^\infty - \frac{\mu}{15} \Phi^{(4)}(x) \Big|_{-\infty}^\infty + \\
&& \hspace{1.2cm}+ \mu \int_{-\infty}^\infty  dx \Big[ \frac{2}{3} \Phi'(x)^3 +\frac{8}{3} \Phi(x) \Phi'(x) \Phi''(x) - \frac{4}{3} \Phi(x) \Phi'(x) \Phi''(x) \Big] =\\
& &\hspace{0.8cm} = -\frac{32}{15} \, \mu \, \nu^5 + \mu\int_{-\infty}^\infty dx \Big[ \frac{2}{3} \Phi'(x)^3 -\frac{2}{3} ( \Phi'(x))^3 \Big]= -\frac{32}{15} \, \mu \, \nu^5
\end{eqnarray*}
where we have used that $\frac{\partial}{\partial x} [\Phi(x) (\Phi'(x))^2] = (\Phi'(x))^3 + 2 \Phi(x) \Phi'(x)\Phi''(x)$ and therefore
\[
\int_{-\infty}^\infty dx \Phi(x) \Phi'(x)\Phi''(x) = \frac{1}{2} \Phi(x) (\Phi'(x))^2 \Big|_{-\infty}^\infty - \frac{1}{2} \int_{-\infty}^\infty dx (\Phi'(x))^3
\]
These results lead to the general formula
and we obtain for the Fermi number:
\begin{equation}
N(H) = -\frac{1}{8\pi} \sum_{n=2}^\infty \frac{2^{n+1}(n-2)!}{(2n-3)!!} \frac{\mu\, \nu^{2n-3}}{(\nu^2+\mu^2)^{n-1}} \label{fermionicnumber03}
\end{equation}
In these computations a very important, but expected point, must be stressed: {\textit{The Seeley coefficients are independent of the specific dependence on $x$ of the Kink profile $\Phi$.}}

Denoting the partial sum of the series (\ref{fermionicnumber03}) as
\[
N^{(n_{\rm max})}= \frac{1}{2} \sum_{n=2}^{n_{\rm max}} \frac{2^{n+1}(n-2)!}{(2n-3)!!} \frac{\mu\, \nu^{2n-3}}{(\nu^2+\mu^2)^{n-1}}
\]
we find that this partial sum may be written in terms of special functions as:
\[
N^{(n_{\rm max})}= - \frac{\mu}{2\sqrt{\mu^2}} +\frac{1}{\pi} \arctan \frac{\mu}{\nu} + \frac{\mu \,\nu^{2n_{\rm max}-1}\Gamma(n_{\rm max})\,{}_2F_1[1,n_{\rm max},\frac{1}{2}+n_{\rm max},\frac{\nu^2}{\nu^2+\mu^2}]}{2\sqrt{\pi} \, (\nu^2+ \mu^2)^{n_{\rm max}} \, \Gamma[\frac{1}{2}+n_{\rm max}]}
\]
We then plot these partial sums
\[
N^{(n_{\rm max})}(z)= - \frac{z}{2\sqrt{z^2}} +\frac{1}{\pi} \arctan z +\frac{\Gamma(n_{\rm max})}{2\sqrt{\pi} \,  \Gamma[\frac{1}{2}+n_{\rm max}]}  \frac{z\,\, {}_2F_1[1,n_{\rm max},\frac{1}{2}+n_{\rm max},\frac{1}{1+z^2}]}{(1+z^2)^{n_{\rm max}}}
\]
as functions of $z=\frac{\mu}{\nu}=\frac{m}{ga}$, see Figure 1.

\begin{figure}[h]
\centerline{\includegraphics[height=4.5cm]{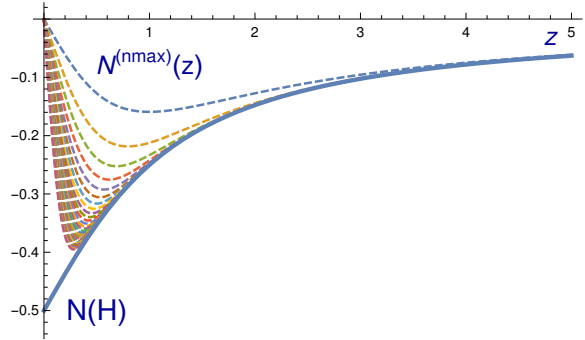}}
\caption{Partial Sums plotted as functions of $\mu/\nu$ for increasing values of $n_{\rm max}$}
\end{figure}

This Figure allows us to grasp an analytical understanding of the application of the Gilkey-de Witt asymptotic method to this kind of calculations. The behavior of $N^{(n_{\rm max})}$ is displayed in Figure 1. We observe that when $\mu\neq 0$ the partial sums converge to the exact response (shown as the lower blue solid line in the Figure). For small values of $m$ this convergence is very slow. For $\mu=0$ (which involves the presence of zero modes in the operator $H$) the response of the formula (\ref{fermionicnumber03}) is zero in contrast with the exact value $N=\pm \frac{1}{2}$. We can impose that when zero modes are present the exact response can be obtained by demanding the continuity of $N(H)$ as a function of $\mu$ at $\mu=0$.

Let us stress some features the method presented above:
\begin{enumerate}
\item This method singles out a value of the mass gap parameter that is consistent with global properties of the operator and that ensures convergence of the perturbation series.
\item The terms that improve the large $\bbeta$ behavior of the heat kernel, see (\ref{htae}) could have been included in (\ref{hkhk}). As one can easily see, such terms vanish upon variation with respect to $\rho$. This observation  explains \textit{a posteriori} why the expansion in Sections \ref{sec:hk} and \ref{GW} worked well without any improvement.
\item This method is actually quite general. An extension to the abelian Goldstone-Wilczek model is quite straightforward. (We do not present it here). In other generalizations one depends just on the computer power.
\end{enumerate}

\section{Conclusions}
In this paper, we suggested a method to compute the fermion number of solitons based on the heat kernel expansion. A crucial step was an introduction of a localizing function $\rho$ in the $\eta$ function that allowed us to use the standard heat kernel coefficients and simplified combinatorics of the problem. With the use of this method we obtain an expression for the fermion number in a multiflavor extension of the Goldstone-Wilczek model.

This method, being just a more systematic version of the usual derivative expansion, shares some drawbacks of the latter. It does not take into account the large distance and short proper time behavior of the heat kernel. Thus, it is not easy to explore the convergence. However, there is an improvement of the expansion \cite{Alonso:2012} which allows to address this problem. With the help of this improved expansion we first reconfirmed the Niemi-Semenoff formula and then demonstrated the convergence of heat kernel series for fermion number.

We would like to stress, that the methods presented here are extendable to more complicated and higher-dimensional models. We are going to consider these models in a future work.

\subsection*{Acknowledgments}
This work was supported in parts by the S\~ao Paulo Research Foundation (FAPESP), projects 2016/03319-6 and 2017/50294-1 (SPRINT), by the grants 303807/2016-4 and 428951/2018-0 of CNPq, by the RFBR project 18-02-00149-a and by the Tomsk State University Competitiveness Improvement Program.
AAI and JMG also thank to the JCyL for partially supporting their research under Grant BU 229P18

\end{document}